\documentclass[3p]{elsarticle}

\usepackage{comment}
\usepackage{amsmath}
\usepackage{latexsym}
\journal{Journal of Nuclear Materials}
\begin{document}
\begin{frontmatter}

\title{Model for screening of resonant magnetic perturbations by plasma in a realistic tokamak geometry and its impact on divertor strike points}
\author[ipp]{P. Cahyna\corref{cor1}\fnref{pres}}
\ead{cahyna@ipp.cas.cz}
\author[cea]{E. Nardon}
\author{JET EFDA Contributors\fnref{jet}}
\address[ipp]{Institute of Plasma Physics, AS CR, v.v.i., Association EURATOM/IPP.CR, Za Slovankou 3, 182 00 Prague, Czech Republic}
\address[cea]{Association EURATOM-CEA, CEA/DSM/IRFM, CEA Cadarache,
  13108 St-Paul-lez-Durance, France}
\fntext[pres]{Presenting author}
\fntext[jet]{See the Appendix of F. Romanelli et al., Proceedings of the 22nd IAEA Fusion Energy Conference 2008, Geneva, Switzerland}
\cortext[cor1]{Corresponding author}

\begin{abstract}
This work addresses the question of the relation between strike-point
splitting and magnetic stochasticity at the edge of a poloidally
diverted tokamak in the presence of externally imposed magnetic
perturbations. More specifically, ad-hoc helical current sheets are
introduced in order to mimic a hypothetical screening of the external
resonant magnetic perturbations by the plasma. These current sheets,
which suppress magnetic islands, are found to reduce the amount of
splitting expected at the target, which suggests that screening
effects should be observable experimentally. Multiple screening
current sheets reinforce each other, i.e. less current relative to the
case of only one current sheet is required to screen the perturbation.
\end{abstract}

\begin{keyword}
Theory and Modelling \sep Plasma properties \sep Plasma-Materials Interaction
\end{keyword}

\end{frontmatter}

\newpage 
\section{Introduction}

An axisymmetric, single null, poloidally diverted tokamak has two
strike-lines where the plasma hits the divertor targets: one on the
high field side (HFS) and one on the low field side (LFS). In the
presence of a non-axisymmetric magnetic perturbation, these
strike-lines are replaced by spiralling patterns \cite{pomphrey:938}. If an
experimental profile (e.g. D$_{\alpha }$ or infrared [IR]) is taken
along the radial direction at a given toroidal location, the
strike-points are then observed to split. Such a splitting is for
instance commonly observed during locked modes~\cite{Evans1995235}. In
the presence of external resonant magnetic perturbations (RMPs) a
splitting may also be expected. On DIII-D, the splitting is observed
during Edge Localized Mode (ELM)-suppressed discharges using $n=3$
perturbations from the I-coils
\cite{0741-3335-50-12-124029,Evans2007570}. It is important to notice
that the splitting is seen much more clearly on particle flux
(D$_{\alpha }$) profiles than on heat flux (IR) profiles, at least in
low collisionality experiments~\cite{0741-3335-50-12-124029} (at high
collisionality, the splitting is however clearly observed on the heat
flux~\cite{Evans2007570}). Recently, DIII-D also reported on splitting
observations (both on heat and particle fluxes) in L-mode plasmas~\cite{SchmitzPSI2010inpress}. JET (using the Error Field Correction Coils (EFCCs)) and MAST
(using $n=3$ perturbations from the ELM control coils~\cite{ericeps09}) find
consistent effects on the heat flux profiles: the splitting is
observed in L-mode but not in H-mode~\cite{NardonPSI2010inpress}.

In the field of ELM control by RMPs from perturbation coils, one major question is to know whether the RMPs stochastize the magnetic field at the edge of the plasma, as assumed by the vacuum modelling. Studies based on the vacuum field assumption~\cite{fenstermacher:056122} have led to a design criterion for the considered ITER ELM control coils~\cite{0029-5515-49-6-065012}. However, two important elements cast doubt on the stochastization of the magnetic field. The first one is the absence of a degradation of the electron temperature gradient in the edge transport barrier, which would be expected in the presence of a stochastic field~\cite{PhysRevLett.40.38}. The second one is the strong rotational screening effect~\cite{fitzpatricknf,fitzpatrickpop} found in simulations of the DIII-D ELM suppression experiments~\cite{heyn2008,ericmodel}. On the other hand the RMPs can also become amplified by the interaction with MHD modes~\cite{park:056115} --- an effect which works against rotational screening~\cite{0029-5515-48-11-115004}.

In this paper, we analyse the possible consequences of the rotational screening (without taking into account the possibility of amplification by MHD modes) on the strike-point splitting in order to assess whether screening effects may explain the absence of a clear splitting of the heat flux profiles in some experiments, in particular in the DIII-D, JET and MAST H-mode discharges referred to above.

\section{Modelling and theoretical understanding of the strike-point splitting}

Under a non-axisymmetric perturbation the magnetic separatrix splits into two surfaces: the stable and unstable manifolds of the X-point. The stable (resp. unstable) manifold is the set of field lines that asymptotically approach the X-point when followed in the direction of (resp. opposite to) the magnetic field. The manifolds are of interest to experiments because they delimit the first passage through the wall of field lines arriving from the plasma core. Their intersections with the divertor plates thus define areas (divertor footprints) where high heat and particle fluxes are carried from the plasma core along the field lines~\cite{Evans05}. Those areas take typically the form of spirals of high temperature and particle recycling around the original (unperturbed) divertor strike point. 

The divertor footprints can be visualized by plotting a map of the
connection length on the divertor plates (a laminar plot)~\cite{WingenPoP09}. The connection length is the distance (measured as the number of toroidal turns) needed to reach the wall again by following a field line starting at a given position. Field lines with large connection lengths remain in the plasma for many turns and carry high fluxes from the hot plasma core. The extent of the footprint can be approximated analytically using the Melnikov function~\cite{wiggins} whose maximum is the difference of $\psi$ between the unperturbed strike point and the tip of the footprint~\cite{pavel-eric-footprints}. When the perturbation has one dominant toroidal mode further simplification is possible and the difference of $\psi$ can be expressed using a single number --- the one-mode Melnikov integral $\tilde {M}_n$~\cite{pavel-eric-footprints}.

An example of the laminar plot and the stable manifold is shown on Figure~\ref{fig:laminar} (left plot) for an equilibrium predicted for the COMPASS tokamak~\cite{Panek2006} in the case of a magnetic field of 1.2~T, low triangularity, single-null (SND) geometry and heating by one co-injected neutral beam~\cite{0741-3335-52-4-045008}. The $n=2$ perturbation is imposed by the existing perturbation coils whose description can be found in~\cite{0029-5515-49-5-055024}.

\section{Physics-motivated method for taking screening currents into account}

\subsection{Coordinate system and resonant field components}

We use an $\left( {s,\varphi ,\theta ^\ast } \right)$ system of
equilibrium coordinates, where $s \mathrel\equiv | ( \psi -\psi _{\text{axis}}) \mathbin/ ( {\psi _{\text{sep}} -\psi _{\text{axis}} }) |^{1/2}$ (with $\psi $ the poloidal magnetic flux), $\varphi $ the geometric toroidal angle and $\theta ^\ast $ the corresponding straight field line poloidal angle (i.e. such that $\left| {{d\varphi } / {d\theta ^\ast }} \right|=\mbox{const.}=q$ along a field line, with $q$ the safety factor). 

Magnetic islands are known to arise from the component of the magnetic perturbation which is perpendicular to the equilibrium flux surfaces. We characterize the latter by the quantity $b^1\equiv {\vec {B}\cdot \vec {\nabla }s} / {\vec {B}\cdot \vec {\nabla }\varphi }$. It can be shown that its Fourier components $b_{mn}^1 $ are directly related to the half-width of the magnetic islands and moreover they are related to the Melnikov integral~\cite{joseph2008}, more precisely proportional to a function which is a generalization of the one-mode Melnikov integral $\tilde {M}_n $ (the Melnikov-like function)~\cite{pavel-eric-footprints} or equivalently the Poincar\'{e} integral~\cite{abdullaev:042508}. At the same time the radial extent of the invariant manifolds (estimated by the Melnikov integral) gives the lower bound of the stochastic layer width because the intersections of the invariant manifolds (the homoclinic tangle) create themselves a thin stochastic layer. The actual stochastic layer can be much wider than this lower bound because it is formed also by the overlap of the magnetic islands~\cite{joseph2008}.

\subsection{Model of the screening currents}

Without loss of generality, we consider one toroidal mode $n$ of the screening currents in the plasma. A generic current can be represented a Fourier sum of these modes which are independent due to the toroidal symmetry.

Screening currents are modelled under the following assumptions:
\begin{enumerate}\item 
  They are radially localized on infinitesimally thin layers around
  the resonant surfaces: $\vec {\jmath}=\sum\nolimits_{q\in S} \delta
    ( s-s_{q,n} )\vec {\jmath}_{q,n} $, where $\delta $ is
  the Dirac delta function and $S$ is the set of rational values of
  the safety factor which define the screening surfaces: $q=m/n$ for
  integer $m$ and the given toroidal mode number $n$. The
  corresponding values of the radial coordinate $s$ are noted as
  $s_{q,n} $.

\item They are parallel to the equilibrium field lines: $\vec
  {\jmath}_{q,n} ={j_{q,n} } / {B_\text{eq} }\cdot \vec {B}_\text{eq} $.

\item They are divergence-free: $\vec {\nabla }\cdot \vec {\jmath}=0$,
  which implies that $\alpha _{q,n} \equiv {j_{q,n} }/{B_\text{eq} }$
  is constant on a field line.
\end{enumerate}

The first assumption corresponds to the fact that the current density is generally localized in a thin layer around the resonant surface \cite{fitzpatricknf,fitzpatrickpop,hazeltinemeiss}. The second assumption follows from the fact that the screening currents are induced to oppose the radial perturbation and to create a radial screening field perpendicular to the field lines a parallel current is needed. The third assumption expresses quasi-neutrality. 

The angular dependence of $\alpha_{q,n} $ has the form of one Fourier mode:
\begin{equation}
\label{eq1}
\alpha _{q,n} \left( {\theta ^\ast ,\varphi } \right)=\Re \left( {\beta _{mn} \exp \left[ {\mbox{i}\left( {m\theta ^\ast +n\varphi } \right)} \right]} \right)
\end{equation}
with $m=qn$, where the toroidal dependence is the consequence of
working with one toroidal mode, while the poloidal dependence follows
from the third requirement: $\alpha _{q,n} $ constant on field
lines. $\beta _{mn} $ is a complex quantity containing both the
amplitude and phase of $\alpha _{q,n} $. Thus, the screening current density can be expressed as a linear combination of basis currents $\vec {\jmath}_{mn,0} $
\begin{equation}
\label{eq2}
\vec {\jmath}=\Re \left( {\sum\limits_{m;m/n\in S} {I_{mn} \vec {\jmath}_{mn,0} } } \right),
\end{equation}
with coefficients $I_{mn} \equiv {\beta _{mn} B_\text{ref} }/ {j_\text{ref} }$ and the basis currents
\begin{equation}
\label{eq3}
\vec {\jmath}_{mn,0} \left( {s,\theta ^\ast ,\varphi } \right)\equiv \frac{j_\text{ref} }{B_\text{ref} }\delta \left( {s-s_{q,n} } \right)\exp \left[ {\mbox{i}\left( {m\theta ^\ast +n\varphi } \right)} \right]\cdot \vec {B}_\text{eq} .
\end{equation}
${B_\text{ref} } / {j_\text{ref} }$ is an arbitrary value expressing the choice of basis current amplitudes (and thus the normalization of $I_{mn})$ relative to the magnetic field strength.

\subsection{Coupling matrix}

We calculate the field $\vec {B}_{mn,0} $ created by the base resonant
current $\vec {\jmath}_{mn,0}$. The numerical method approximates $\vec
{\jmath}_{mn,0}$ by discrete helical current filaments on the
screening surface and calculates their vector potential on a flux
surface aligned mesh. It avoids the mesh points which are too close to
the screening surface and uses instead a cubic spline interpolation in order that the discrete approximation of currents not cause an error. $\vec {B}_{mn,0} $ is then obtained as the curl of the interpolated vector potential, which gives all its components and automatically satisfies the condition of zero divergence. The corresponding $b^1$ component, denoted $b_0^{1,mn} $, is then Fourier transformed at each resonant surface $q={m'}/ n'$ in order to obtain the resonant components $b_{{m}'{n}',0}^{1,mn} $. Thus, $b_{{m}'{n}',0}^{1,mn} $ designates the resonant part, on the $q={m'} / n'$ surface, of the $b^1$ created by a resonant current $\vec {\jmath}_{mn,0}$ located at the $q=m/ n$ surface. Due to the toroidal symmetry of the field equations in $b_{{m}'{n}',0}^{1,mn} $ we have $n={n}'$, otherwise $b_{{m}'{n}',0}^{1,mn} =0$.

The plasma response field corresponding to the total current $\vec
{\jmath}$ as given by Eq. (\ref{eq2}) is $\vec {B}_{\text{plasma}} =\Re
\left( \sum_{m;m/n\in S} I_{mn} \vec {B}_{mn,0} 
\right)$ whose resonant $b^1$ components are $b_{m'n,\text{plasma}}^1
=\sum\nolimits_m b_{{m}'n,0}^{1,mn} \cdot I_{mn}  / 2 $ and $b_{-m'-n,\text{plasma}}^1 =\sum\nolimits_m \left( b_{{m}'n,0}^{1,mn} \cdot I_{mn}  \right)^\ast  / 2 $. The RHS is the product of the matrix $b_{{m}'n,0}^{1,mn} $ with subscripts $m$ and ${m}'$, which we call the coupling matrix, by the current vector $I_{mn} $.

\subsection{Calculation of the screened field by inversion of the coupling matrix}

To determine the coefficients $I_{mn} $ one needs an assumption about the character of the plasma response, i.e. if it amplifies or screens the perturbation and by what amount. In the following we assume an efficient screening which completely eliminates magnetic islands at the rational surfaces in the pedestal region, i.e. the resonant Fourier components $b_{mn}^1 $ of the total magnetic field are zero.

The procedure to obtain the screened field begins with the calculation of the coupling matrix, for a given choice of the set of screening surfaces $S$. Independently, the vacuum RMP spectrum $b_{m'{n}',\text{vac}}^1 $ is calculated from the coil geometry. The screening current distribution $I_{mn}^{\text{screen}} $ is obtained by solving $\sum\nolimits_m {b_{m'n,0}^{1,mn} \cdot I_{mn}^{\text{screen}} } =-2b_{m'n,\text{vac}}^1 $, i.e. by inverting the coupling matrix. The full, screened field $\vec {B}_{\text{full}} $ is then obtained as $\vec {B}_{\text{full}} =\vec {B}_{\text{vac}} +\Re \left( {\sum\nolimits_m {I_{mn}^{\text{screen}} \vec {B}_{mn,0}} } \right)$. It is easy to verify that its resonant Fourier components $b_{mn,\text{full}}^1$ on rational surfaces with $q=m / n\in S$ satisfy the property $b_{mn,\text{full}}^1 =0$ up to the error introduced by the numerical method.

If the set $S$ consists of only one screening surface, the coupling
matrix is trivial, with one element. The screening currents for each
surface alone are given by using only the diagonal terms of the
coupling matrix: $I_{mn,\text{diag}}^{\text{screen}} =-2{b_{mn,\text{vac}}^1
} / {b_{mn,0}^{1,mn} }$. If
more screening surfaces are considered, the currents $I_{mn}^{\text{screen}}
$ on each one required to cancel the resonant components may be
different than for one surface alone due to off-diagonal components of
the coupling matrix. This effect can be quantified by the ratio
between $I_{mn}^{\text{screen}} $ calculated using the full matrix and
$I_{mn,\text{diag}}^{\text{screen}} $. If for example the ratio is lower than
1, it means that the perturbation field from different screening
surfaces reinforce each other and smaller currents are required for
screening than if screening were due only to one surface. This is the
case for all the examples described in the next subsection, where we
present actual values of this ratio. This is one of the geometry
effects neglected in cylindrical models such as
\cite{heyn2008,ericmodel,marinaiaea2008nf} where the coupling matrix is
always diagonal.
\begin{figure*}%
  \centering%
  \includegraphics[width=75mm]{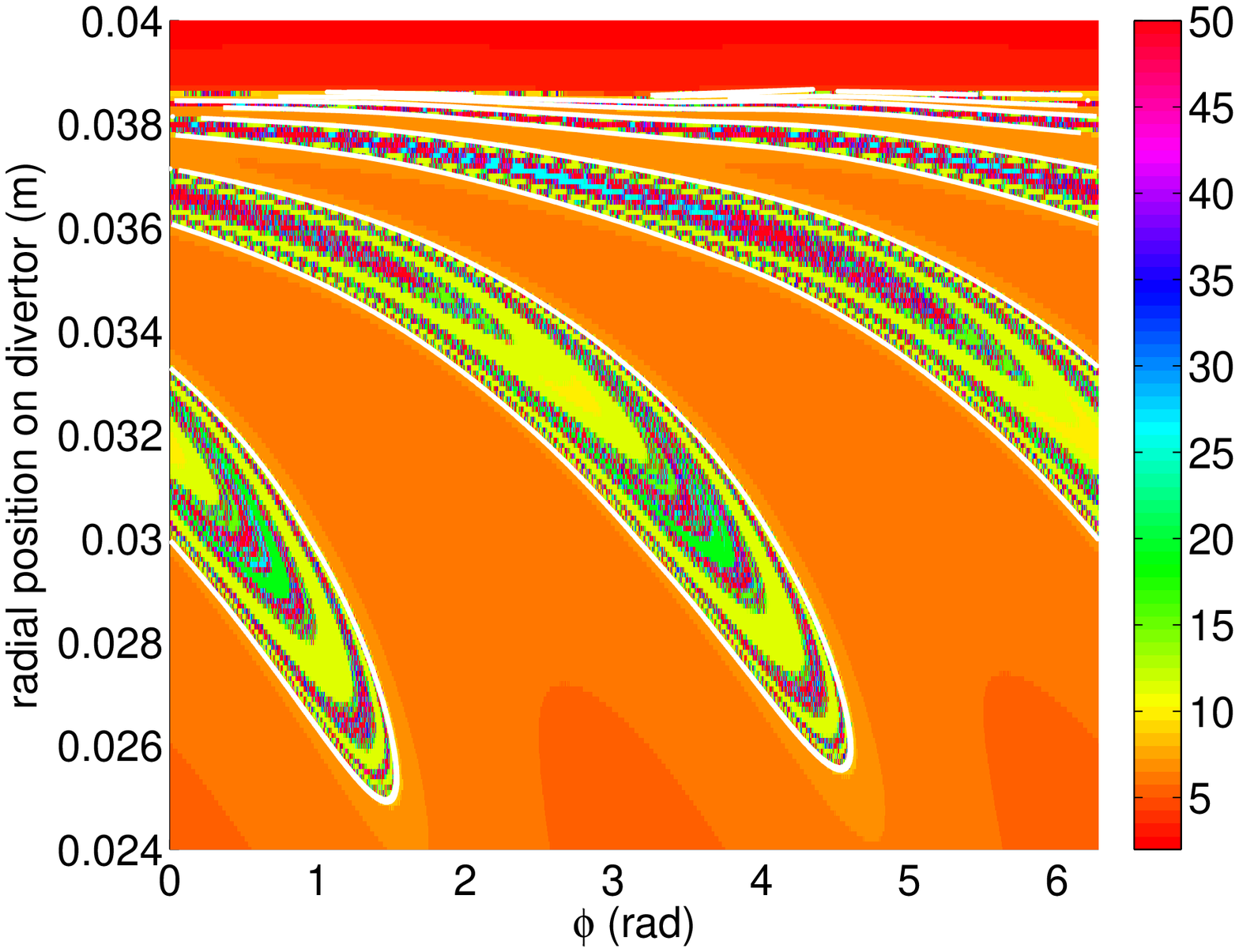}%
  \includegraphics[width=75mm]{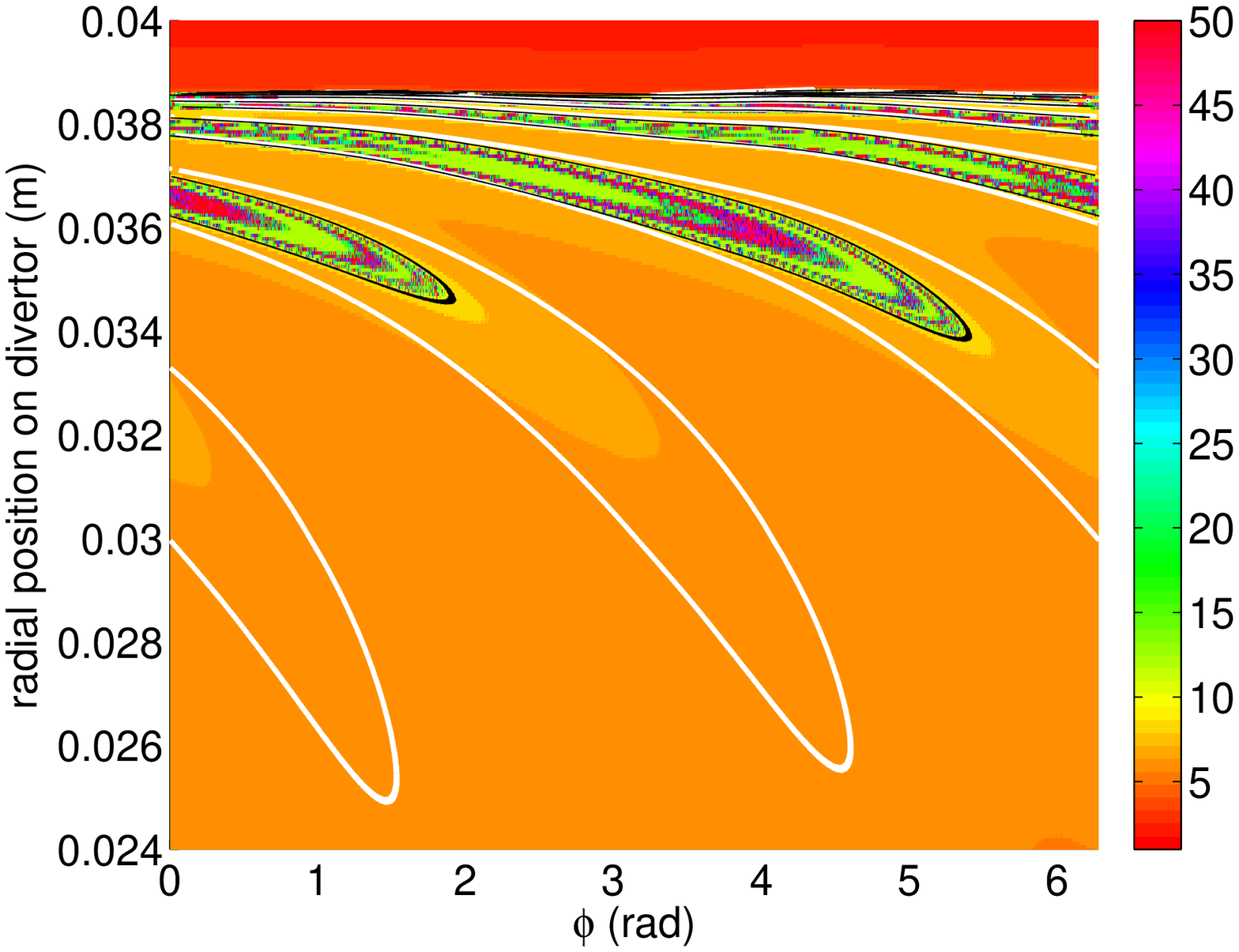}%
  \caption{Laminar plot of the
  connection length (as a number of toroidal turns) on the divertor of
  COMPASS near the HFS strike point with a vacuum (left) and screened
  (right) perturbation field. White line: The stable manifold of the
  vacuum perturbation field. Black line: the stable manifold of the
  screened perturbation field.}
\label{fig:laminar}
\end{figure*}

\subsection{Numerical examples}

\begin{figure}[htb]%
  \centering%
  \includegraphics[width=75mm]{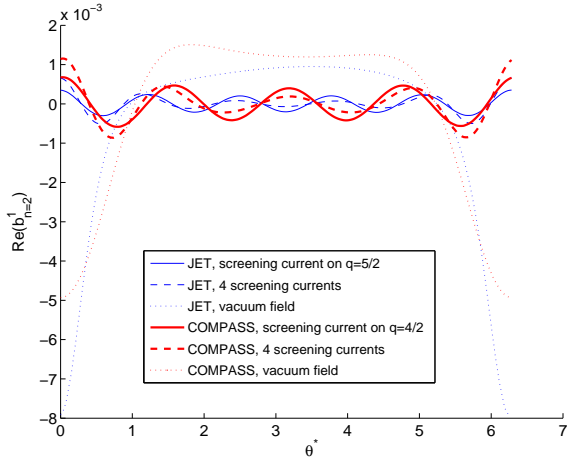}%
  \caption{Poloidal dependence of the real part of the $n=2$ component
  of the relative perturbation field $b^{1}$ at the $q=5/2$ resonant
  surface for JET, shot {\#}79729 and $q=4/2$ for COMPASS. Fields of one
  screening current (full lines) and four screening currents (dashed
  lines) are shown.}
\label{fig:field}
\end{figure}

We chose two cases to illustrate the method and to show the effect of
screening on footprints: the COMPASS case described above and an
equilibrium from an ELM control experiment on JET with $n=2$
perturbation of the EFCCs~\cite{NardonPSI2010inpress} (shot {\#}79729
at 19.38s). For both cases we first calculate the screening field
needed to cancel the perturbation on a single surface ($q=4/ 2$ for
COMPASS, $q=5/ 2$ for JET). In those cases the coupling matrix is
trivial, with one element. Then we calculate the screening field
choosing four screening surfaces with $q=4 / 2,\ldots,7 / 2$ for
COMPASS, $q=5/2,\ldots,8/2$ for JET. The $n=2$ mode of the screening
field $b_{\text{screen}}^1 $ as a function of $\theta ^\ast $ on the
innermost screening surface is shown on Figure~\ref{fig:field} for the
four cases (each equilibrium with one and four screening
surfaces). The field is given on innermost surface in order to show how the
field changes depending on the choice of screening currents while
maintaining the same resonant component. The screening field of one screening surface is distributed all over the resonant surface, due to the helical structure of the screening current, while for four screening surfaces the screening field is mostly localized on the LFS. Both COMPASS and JET show this effect. It shall be noted that the vacuum field is also localized at the LFS because of the position of the coils in both tokamaks, so the screening field of four currents is more similar to the vacuum field than the field of a single current. The ratio $I_{m=5\,n=2}^{\text{screen}} /I_{m=5\,n=2,\text{diag}}^{\text{screen}} $ for JET is 0.63, for COMPASS $I_{m=4\,n=2}^{\text{screen}} /I_{m=4\,n=2,\text{diag}}^{\text{screen}}$ is 0.72, with similar and generally decreasing values for other surfaces.

The numerical error in determining the screening current amplitudes
was estimated by increasing four times the resolution of the
discretization of the screening currents (from 4096 to 16384). The largest change was observed for the outermost surface (6.3{\%} for COMPASS, 8.3{\%} for JET). The error decreases for the inner surfaces down to 0.1{\%} for COMPASS, 0.2{\%} for JET at the innermost surface.

\section{Impact of screening currents on the splitting}

\begin{figure}%
  \centering%
  \includegraphics[width=75mm]{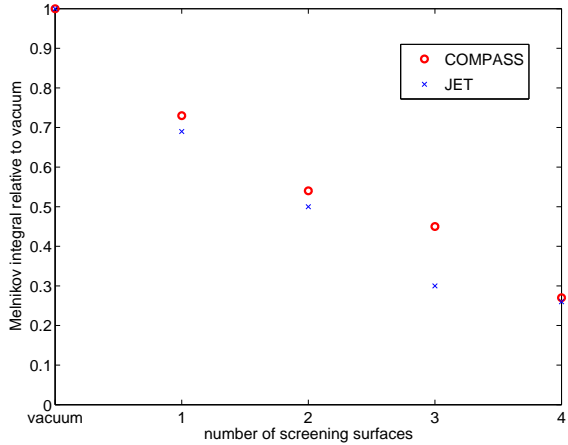}%
 \caption{The relative Melnikov integral from Table 1
  as a function of the number of screening surfaces: from one at the
  $q=5/2$ resonant surface for JET, shot {\#}79729 and $q=4/2$ for
  COMPASS, up to four of them.}
\label{fig:melnikov}
\end{figure}

For the $b_0^{1,mn} $ field created by the resonant current $\vec {\jmath}_{mn,0}$ we may compute the one-mode Melnikov integral at the separatrix which we will note $\tilde {M}_0^{mn} $. The total Melnikov integral which estimates the splitting from all the screening currents and the vacuum field is
\begin{equation}
\label{eq4}
\tilde {M}_n =\tilde {M}_{n,\text{vac}} +\sum\limits_{m;m/n\in S} {I_{mn}^{\text{screen}} \tilde {M}_0^{mn} } 
\end{equation}
The ratio of the footprint extent measured in terms of $\psi$ for the screened vs. the vacuum field is given by $\left| {\tilde {M}_n } \right|/\left| {\tilde {M}_{n,\text{vac}} } \right|$. Table~\ref{tab1} lists this value for the COMPASS and JET cases as a function of the choice of screening currents, and Figure~\ref{fig:melnikov} shows the same data.

In both cases a significant reduction of the footprints is predicted
by the Melnikov integral when four screening currents are
considered. To confirm this we plotted the stable manifold and a
laminar plot around the inner strike point for COMPASS for the
screened field and the stable manifold for the vacuum field for
comparison (Figure~\ref{fig:laminar}, right plot). The stable manifold forms the
boundary of the footprints as expected and indeed shows a clear
reduction in comparison to the vacuum stable manifold. The difference
of $\psi$ (normalized to the poloidal flux at the separatrix) between
the footprint tip and the base is $6.8\times 10^{-3}$, while the Melnikov
integral method predicts $5.3\times 10^{-3}$. The inaccuracy of the Melnikov
method is significantly lower (relative error below 1/10) for the
other cases in the table which do not include the outermost ($m=7$)
screening surface. Similar result was found for the JET case (laminar
plots are shown in~\cite{NardonPSI2010inpress}): the actual difference is $3.4\times 10^{-3}$,
while the Melnikov integral method predicts $3.6\times 10^{-3}$. The
outermost ($m=8$) surface can also lead to a large error in the
Melnikov integral estimation, especially with the $m=8$ surface alone,
where the Melnikov integral predicts reduction of footprints by a
factor of 0.89, while the actual reduction is 0.56.
\begin{table}[htbp]
\begin{center}
\begin{tabular}{|l|l|l|l|l|}
\hline
COMPASS & 
$m=4$& 
$m=4,5$& 
$m=4,5,6$& 
$m=4,5,6,7$ \\
\hline
$| {\tilde {M}_2 } |/| {\tilde {M}_{2,\text{vac}} } |$& 
0.73& 
0.54& 
0.45& 
0.27 \\
\hline
JET& 
$m=5$& 
$m=5,6$& 
$m=5,6,7$& 
$m=5,6,7,8$ \\
\hline
$| {\tilde {M}_2 }|/| {\tilde {M}_{2,\text{vac}} }|$& 
0.69& 
0.50& 
0.30& 
0.26 \\
\hline
\end{tabular}
\label{tab1}
\caption{One mode Melnikov integral of the screened perturbation
  normalized to the Melnikov integral of the vacuum perturbation, for
  different choices of the screening currents, given by the resonant
  poloidal mode numbers $m$. $\left| {\tilde
      {M}_{2,vac} } \right|$ is $2.00\times 10^{-2}$ for COMPASS,
  $1.38\times 10^{-2}$ for JET, relative to the poloidal flux at the
  separatrix.}
\end{center}
\end{table}

\section{Discussion and conclusions}

We developed a model of the plasma response currents on resonant
surfaces and the resulting field, based on the realistic geometry of
poloidally diverted tokamak plasmas and thus appropriate for the
region near the separatrix, which is crucial for the ELM mitigation by
external perturbations and also for the impact of perturbations on the
divertor strike points (strike point splitting). To compute the
screening currents we used the assumption of complete screening of
resonant modes in the edge region, because we do not simulate the
plasma response self-consistently. This is justified by earlier
results indicating that the resonant modes of the perturbation will be
suppressed by strong pressure gradient in the pedestal region \cite{heyn2008,ericmodel}. As we do not directly couple our model with these results yet, we performed a scan of different possible combinations of screening currents. In future work we plan to use the results of self-consistent MHD models to determine the set of screening surfaces and the screening factors, which will also allow for incomplete screening. The reason for this approach is that those MHD models use a cylindrical geometry in which the effect on the divertor strike points can't be represented, and our model fills this gap. While there are MHD codes (e.g. NIMROD~\cite{0029-5515-48-11-115004}, M3D~\cite{0029-5515-49-5-055025} and JOREK~\cite{0741-3335-51-12-124012}) using a realistic geometry which thus can model the strike points themselves (as shown for JOREK in~\cite{0741-3335-51-12-124012}) without needing our model, they have limitations in the physics included (lack of realistic resistivity) which justifies the interest of our approach.

The resulting screened field was used to model magnetic footprints on
the divertor by tracing the field lines. For two example cases
(single-null equilibria of COMPASS and JET with $n=2$ perturbations)
we have shown notable differences in comparison with the vacuum
field. The screening significantly reduces the spiralling patterns of
field lines coming from inside the plasma. The spirals are shortened
along their axis, the position of the axis is not affected. Reducing
the coil current in a vacuum model has a similar impact. Comparison
with experimental observation of strike point splitting could be thus
used for validating the starting assumption about the
screening. Coupling with MHD models will also enable us to do scans of
the dependence of strike point splitting on the collisionality (and
thus resistivity) and rotation, which are both important parameters
for the screening effect, and compare the results with experiments. It
can be expected that higher collisionality and slower rotation will
reduce the screening and enhance the footprints. The discussion of
experimental results is however more complicated than a simple
comparison with the predicted magnetic footprints because of the
deformation of flux surfaces beyond the separatrix. The particles in
the scrape-off layer would follow those distorted surfaces, and the
screening may change their distortion differently than in the
case of the separatrix. This shows that it is necessary to distinguish
between particle and heat flux and to evaluate the impact of screening
on flux surface distortion in the SOL in addition to the separatrix. Indeed, as mentioned in the Introduction, in the DIII-D low collisionality experiments the splitting of particle flux and heat flux are different. Transport modelling using the vacuum perturbation failed to explain this observation~\cite{0029-5515-50-3-034004} and it will be thus interesting to repeat it for the screened field as calculated by our model to see if the screening can provide an explanation.

The reduction of footprints can be efficiently estimated by the Melnikov integral method, but it is sometimes inaccurate when a screening current very close to the separatrix is included, probably because of the strong variation of the screening field while the Melnikov method (a first-order perturbation method) needs the perturbation field to be slowly varying in the vicinity of the separatrix.

The reduction of footprints can be qualitatively understood from the
fact that the screening field is mostly localized at the LFS, just as
the field of the coils. It can be shown that for a LFS-localized
perturbation field the Melnikov integral which estimates the splitting
is linked to the values of the Melnikov-like function on the resonant
surfaces which is proportional to the resonant modes of the
perturbation~\cite{pavel-eric-footprints}. Thus eliminating the
resonant modes shall also mostly eliminate the Melnikov integral and
splitting. It shall be noted that for a single screening surface the
field is not at all localized so this reasoning does not
apply. Indeed, we have seen that a single screening surface reduces
the Melnikov integral only weakly. Multiple screening surfaces also
require less current to screen the resonant component of the field
than a single screening surface ---
we may say that the currents reinforce each other. This effect is
due to the realistic geometry used here and can't be represented in the
cylindrical approximation used in some models for numerical simplicity.

Our model is designed to represent one significant feature of the plasma response to the perturbations --- the surface screening currents localized at resonant surfaces. The ideal MHD models assume them so the opening of magnetic islands is prevented~\cite{park:056115}, while the resistive MHD models predict them self-consistently. There are however other ways of plasma reaction to the perturbation, namely coupling to MHD modes which can provide amplification of the applied perturbation \cite{park:056115,0029-5515-48-11-115004,lanctot:030701}, especially important at high $\beta$~\cite{0029-5515-49-11-115001}, and those are not represented in our approach.

\section{Acknowledgements}

We thank V. Fuchs for providing the COMPASS magnetic equilibrium used
in the simulations. The access to the MetaCentrum computing facilities
provided under the research intent MSM6383917201 is appreciated. This
work, supported by AS CR {\#}AV0Z20430508, MSMT CR {\#}7G09042 and {\#}LA08048 and by the European Communities under the contracts of Association between EURATOM and IPP.CR and CEA, was carried out within the framework of the European Fusion Development Agreement. The views and opinions expressed herein do not necessarily reflect those of the European Commission.

\bibliography{Cahyna-Nardon-PSI-JNM}
\bibliographystyle{model1a-num-names-url}

\end{document}